\documentclass[english,a4paper]{article}
\usepackage[T1]{fontenc}
\usepackage[cp1250]{inputenc}
\usepackage{amssymb}
\usepackage{hyperref}
\usepackage{amsfonts}
\usepackage{graphicx}

\begin{document}

\title{Singular extremals for the time-optimal control of dissipative spin 1/2 particles}

\author{M. Lapert, Y. Zhang, M. Braun, S. J. Glaser\footnote{Department of Chemistry, Technische Universit\"at M\"unchen, Lichtenbergstrasse 4, D-85747 Garching, Germany} and D. Sugny\footnote{Laboratoire Interdisciplinaire Carnot de Bourgogne (ICB), UMR 5209 CNRS-Universit\'e de Bourgogne, 9 Av. A. Savary, BP 47 870, F-21078 DIJON Cedex, FRANCE, dominique.sugny@u-bourgogne.fr}}

\maketitle

\begin{abstract}
We consider the time-optimal control by magnetic fields of a spin 1/2
particle in a dissipative environment. This system is used as an
illustrative example to show the role of singular extremals in the
control of quantum systems. We analyze a simple case where the control law is explicitly determined.
We experimentally implement the optimal control using techniques of Nuclear Magnetic Resonance. To our knowledge, this is the first experimental demonstration of singular extremals in quantum systems with bounded control amplitudes.
\end{abstract}

\begin{figure}
\centering
\includegraphics[height=3in]{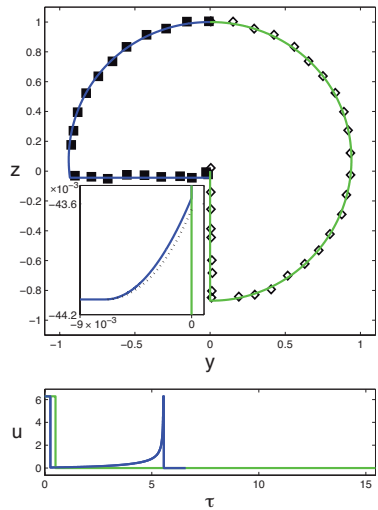}
\caption{\label{fig1} (Color online) Plot of the optimal trajectories (blue
curve) and of the inversion recovery sequence (green curve) in the
plane $(y,z)$ for $T_1=740~\textrm{ms}$, $T_2=60~\textrm{ms}$ and
$\omega_{max}/(2 \pi)=32.3$ Hz. The experimentally measured
trajectories are represented by filled squares and open diamonds,
respectively. The corresponding control laws are represented in
the lower panel. In the upper panel, the small insert represents a
zoom of the optimal trajectory near the origin. The dotted line is
the switching curve originating from the horizontal singular line.
The vertical green line corresponds to the intuitive solution. The
solid blue curve is the optimal trajectory near the origin.}
\end{figure}

Optimal control theory can be viewed as a generalization of the
classical calculus of variations for problems with dynamical
constraints. Its modern version was born with the Pontryagin
Maximum Principle (PMP) in the late 1950's \cite{pont}. Its
development was originally inspired by problems of space dynamics,
but it is now a key tool to study a large spectrum of applications
such as robotics, economics, and quantum mechanics. Solving an
optimal control problem means finding a particular control law,
the optimal control, such that the corresponding trajectory
satisfies given boundary conditions and minimizes a cost
criterion. Examples of cost functionals of physical interests are
the energy and the duration of the
control. A strategy for solving an optimal control
problem consists in finding extremal trajectories which are
solutions of a generalized Hamiltonian system subject to the
maximization condition of the PMP. In a second step, one selects
among the extremals the ones which effectively minimize the cost
criterion. Although its implementation looks straightforward, the
practical use of the PMP is far from being trivial and each
control has to be analyzed using geometric and numerical methods.
The first applications of optimal control theory in quantum dynamics began in the mid 80's \cite{conolly}. Continuous
advances have been done both theoretically and experimentally
\cite{rice}. Among the set of extremal trajectories determined from the PMP, one distinguishes regular and singular ones (see below for a concrete definition). Surprisingly, whereas regular extremals are well-known, the existence and the
potential applications of singular extremals in quantum control have been largely ignored
in the chemical-physics literature and only few results exist
\cite{alessandro,khaneja,boscain,boscain2,BS,rabitz}. Note that
the zero field used in the optimal control of quantum systems with
unbounded control \cite{khaneja,tannor} can be viewed as a limit
case of singular extremals when the maximum amplitude of the
control goes to infinity (see \cite{boscain} for details). In this
paper, we consider a simple physical example, a spin $1/2$
particle in a dissipative environment, which highlights the
crucial role of singular controls. For instance, a gain of 50\% in
the control duration can be obtained over the standard inversion
recovery sequence when using singular extremals (see Fig.
\ref{fig1}). We take advantage of this example to explain the
mathematical framework and to detail in the last section its
physical interpretation. The optimal control law is also
implemented experimentally using techniques of Nuclear Magnetic
Resonance (NMR).

In the considered case of resonant radiation, the state of the
system can be completely represented by a two-dimensional state
vector $X\in \mathbb{R}^2$ and a single control $u$ is sufficient
\cite{BS}. The corresponding controlled system is defined by
differential equations of the form $\dot{X}=F_0(X)+uF_1(X)$ and
the control parameter satisfies $|u|\leq u_0$, which defines the set of admissible controls. The objective of
the control is to determine a function $u(t)$ such that the system
goes in minimum time from the initial point $X_0$ to a target
state $X_1$. We use for that the PMP which can be sketched as follows
\cite{pont,BC,boscainbook}. We introduce the pseudo-Hamiltonian
$\mathcal{H}(X,P,u)=P\cdot (F_0+uF_1)$ where the adjoint state
$P\in\mathbb{R}^{*2}$ for any time $t$. An optimal trajectory is
solution of the equations
\begin{eqnarray*}
\begin{array}{ll}
\dot{X}(t)=\frac{\partial \mathcal{H}}{\partial P}\\
\dot{P}(t)=-\frac{\partial \mathcal{H}}{\partial X}
\end{array}
\end{eqnarray*}
where $u$ is obtained from the maximization condition
$\mathcal{H}(X,P,u)=H(X,P)=\max_{v\in
[-u_0,u_0]}[\mathcal{H}(X,P,v)]$ with the condition $H(X,P)\geq 0$
\cite{boscainbook}. For
controlled systems on the coordinate plane $\mathbb{R}^2$, the solutions of
the PMP take a very simple form. We consider the switching
function $\phi(X,P)=P\cdot F_1$ \cite{boscainbook}, which is the only term of $\mathcal{H}$ on which the control can act. When $\phi(X,P)\neq 0$, one
immediately sees  that the maximization
condition leads to bang controls, i.e. to controls of constant and
maximum amplitude of the form $u=u_0\times \textrm{sign}[\phi]=\pm u_0$. These extremals are the regular ones. If
$\phi$ vanishes in an isolated point then the control field
switches from $\pm u_0$ to $\mp u_0$. The singular situation is
encountered when the switching function vanishes on a given time
interval. In this case, the control cannot be directly determined
by the maximization condition. The control parameter is instead computed by requiring that
$\phi(X,P)=0$ on the singular arc, which leads to the relations
$\phi(X,P)=\dot{\phi}(X,P)=\ddot{\phi}(X,P)=\cdots=0$. This
condition allows to identify the manifolds, here lines of the coordinate
plane on which the singular trajectory lies. Note
also that in this case, the control field is not constantly equal
to $+u_0$ or $-u_0$, but belongs to the interval $[-u_0,u_0]$.
The final optimal control law is determined by gluing together singular and
regular arcs.

One of the most promising fields of applications of quantum control
is the control of spin systems in NMR \cite{spin}. We use such systems in this paper to show the role of
singular extremals in the time-optimal control of quantum
dynamics. In a first step, we consider a spin 1/2 particle in a dissipative environment whose dynamics
is governed by the Bloch equation:
\begin{eqnarray}\label{eq1}
\left(
\begin{array}{l}
\dot{M}_x\\
\dot{M}_y\\
\dot{M}_z
\end{array}\right)=
\left(\begin{array}{l} -M_x/T_2 \\
-M_y/T_2 \\
+(M_0-M_z)/T_1
\end{array}\right)
+\left(\begin{array}{l} \omega_y M_z \\
-\omega_x M_z \\
\omega_x M_y-\omega_y M_x
\end{array}\right)\nonumber
\end{eqnarray}
where $\vec{M}$ is the magnetization vector and
$\vec{M}_0=M_0\vec{e}_z$ is the equilibrium point of the dynamics.
We assume that the control field
$\vec{\omega}=(\omega_x,\omega_y,0)$ satisfies the constraint
$|\vec{\omega}|\leq \omega_{max}$. We introduce the normalized
coordinates $\vec{x}=(x,y,z)=\vec{M}/M_0$, which entails that at
thermal equilibrium the $z$ component of the scaled vector
$\vec{x}$ is by definition $+1$. The normalized control field
which satisfies $|u|\leq 2\pi$ is defined as
$u=(u_x,u_y,0)=2\pi\vec{\omega}/\omega_{max}$, while the
normalized time $\tau$ is given by $\tau=(\omega_{max}/2 \pi) t$.
Dividing the previous system by $\omega_{max}M_0/(2\pi)$, one
deduces that the dynamics of the normalized coordinates is ruled
by the following system of differential equations:
\begin{eqnarray}
\left(
\begin{array}{l}
\dot{x}\\
\dot{y}\\
\dot{z}
\end{array}\right)=
\left(\begin{array}{l} -\Gamma x \\
-\Gamma y \\
\gamma-\gamma z
\end{array}\right)
+\left(\begin{array}{l} u_y z \\
-u_x z \\
u_x y-u_y x
\end{array}\right)
\nonumber
\end{eqnarray}
where $\Gamma=2\pi/(\omega_{max}T_2)$ and $\gamma=2\pi/(\omega_{max}T_1)$.

We consider the control problem of bringing the system from the
equilibrium point $\vec{M}_0$ to the zero-magnetization point
which is the center of the Bloch ball. In the setting of NMR
spectroscopy and imaging, this corresponds to saturating the
signal, e.g. for solvent suppression  or  contrast enhancement,
respectively \cite{Inv_recov}. Since the initial point belongs to
the $z$- axis, it can be shown that the controlled system is
equivalent to a system with only one control where, e.g., $u_y=0$
\cite{BS}. Roughly speaking, this means that the meridian planes of the Bloch sphere
play all the same role for the optimal trajectory. Taking $u_y=0$,
we are thus considering a problem in a plane of the form:
\begin{eqnarray}
\left(
\begin{array}{l}
\dot{y}\\
\dot{z}
\end{array}\right)=
\left(\begin{array}{l} -\Gamma y \\
\gamma-\gamma z
\end{array}\right)
+u\left(\begin{array}{l} -z \\
y
\end{array}\right)
\nonumber
\end{eqnarray}
where the subscript $x$ has been omitted for the control
parameter. We can then apply for this system the theoretical
description of the previous paragraph where $F_0=(-\Gamma
y,\gamma-\gamma z)$ and $F_1=(-z,y)$.

As detailed above, we introduce the switching function
$\phi=P\cdot F_1=-p_y z+p_z y$ \cite{boscainbook}. Using the fact
that $\frac{d\phi}{dt}=P\cdot V$ where $V=(-\gamma +\gamma z
-\Gamma z,-\Gamma y+\gamma y)$ and the relations  $P\cdot
F_1=P\cdot V=0$ on a singular arc, one deduces that the vectors
$F_1$ and $V$ must be parallel on this set since $P$ is non zero.
This means that the singular trajectories belong to the set
$S=\{X\in\mathbb{R}^2 | \textrm{det}(F_1,V)=0\}$ which corresponds
to the union of the vertical line $y=0$ and of the horizontal one
with $z$ given by
 $$z_0=-{{\gamma}\over{2(\Gamma-\gamma)}}=-{{T_2}\over{2(T_1 - T_2)}}$$
  if $\Gamma\neq \gamma$ (or equivalently if $T_1 \ne T_2$). The corresponding singular control $u_s$,
  which is determined from the condition $\ddot{\phi}(X,P)=0$, is given by
\begin{equation}\label{sing}
u_s(y,z)=\frac{-y\gamma(\Gamma-2\gamma)-2yz_0(\gamma^2-\Gamma^2)}{2(\Gamma-\gamma)(y^2-z_0^2)-\gamma z_0}.
\end{equation}
Note that the control $u_s$ is not defined as a function of the time but as a function of $y$ and
$z$. One also deduces that this singular control vanishes on the
vertical singular line and that it is admissible, i.e. $|u_s|\leq
2\pi$, on the horizontal one if $|y|\geq
|\gamma(\gamma-2\Gamma)|/[2\pi(2\Gamma-2\gamma)]$. For smaller
values of $y$, the system cannot follow the horizontal singular
arc and a switching curve appears from the point where the
admissibility is lost \cite{boscainbook}. A switching curve is a
line in the plane $(y,z)$ where the optimal control changes sign
when crossing it.

The optimality of the singular trajectories can be determined
geometrically by introducing the clock form $\alpha$ which is a
1-form such that $\alpha(F_0)=1$ and $\alpha(F_1)=0$. The form
$\alpha$ is defined on points where $F_0$ and $F_1$ are not
collinear. Let $\gamma_1$ and $\gamma_2$ be two extremals starting
and ending at the same points and $\tau_1$ and $\tau_2$ the
corresponding times needed to follow the two trajectories. The
clock form allows to determine the time taken to travel a path
since, for instance,
$\int_{\gamma_1}\alpha=\int_0^{\tau_1}\alpha(\dot{X})d\tau=\int_0^{\tau_1}\alpha(F_0)d\tau=\tau_1$.
To compare $\tau_1$ and $\tau_2$, we consider the loop
$\gamma_1\bigcup \gamma_2^{-1}$ where $\gamma_2^{-1}$ is
$\gamma_2$ run backward. Introducing the surface $D$ delimited by
$\gamma_1$ and $\gamma_2$, a simple computation leads to
$\int_{\gamma_1\bigcup \gamma_2^{-1}}\alpha=\int_Dd\alpha$. Since
$d\alpha$ is equal to zero only on the singular set and remains of
constant sign outside \cite{BC}, one obtains that
$\tau_1-\tau_2=\int_Dd\alpha$. In particular, it can be shown that
the horizontal singular line is locally optimal and that the
vertical one is optimal if $z>z_0$.

The control problem used for illustration is defined by the
relaxation parameters $\gamma^{-1}$ and $\Gamma^{-1}$ (expressed
in the normalized time unit defined above) of 23.9 and 1.94,
respectively. We compare
the optimal control law with an intuitive one used in NMR. The
intuitive solution, the so-called inversion recovery (IR) sequence
\cite{Inv_recov}, is composed of a bang pulse to reach the
opposite point of the initial state along the $z$- axis followed
by a zero control where we let the dissipation act up to the
center of the Bloch ball. The optimal and the IR solutions are
plotted in Fig. \ref{fig1}. Geometric tools allow to show that the
optimal control is the concatenation of a bang pulse, followed
successively by a singular control along the horizontal singular
line, another bang pulse and a zero singular control along the
vertical singular line. Figure \ref{fig1} displays also the
switching curve which has been determined numerically by
considering a series of trajectories with $u=+2\pi$ originating
from the horizontal singular set where $\phi=0$. The points of the
switching curve correspond  to the first point of each trajectory
where the switching function vanishes. We have also checked that
the second bang pulse of the optimal sequence does not cross the
switching curve up to the vertical singular axis. In this example,
a gain of 58\% is obtained for the optimal solution over the
intuitive one, which clearly shows the importance of singular
extremals.
\begin{figure}
\centering
\includegraphics[height=3in]{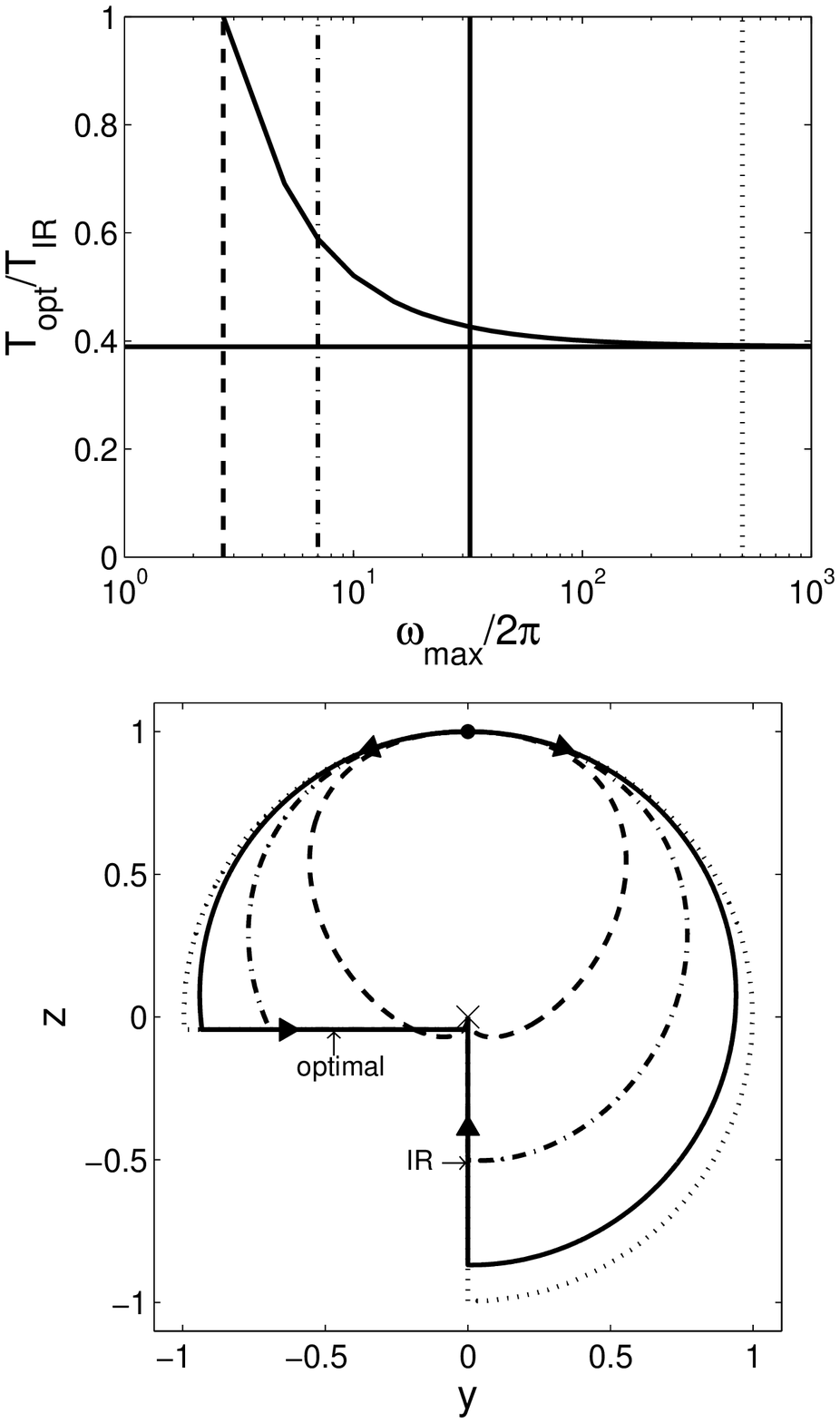}
\caption{\label{fig2} (top) Ratio $T_{opt}/T_{IR}$ as a function
of $\omega_{max}/(2 \pi)$. The horizontal line indicates the
position of the limit ratio when $\omega_{max}\to +\infty$.
(bottom) Optimal trajectories (left part) and the inversion
recovery sequences (right part) for four values of
$\omega_{max}/(2 \pi)$, 2.7, 7, 32.3 and 500 Hz. The vertical lines of the top panel
correspond to the four solutions of the bottom panel. The solid
curve is the case considered in Fig. \ref{fig1}. The black dot and the
cross represent respectively the positions of the initial and
final points. The small arrows indicate the way the trajectories
are followed.}
\end{figure}

Both pulse sequences were implemented experimentally
 on a Bruker Avance 250MHz spectrometer with linearized amplifiers. The experiments
 were performed using the proton spins of H$_2$O. The sample consists of  10\% H$_2$O, 45\%
 D$_2$O and 45\% deuterated Glycerol, saturated with CuSO$_4$. At room temperature (298 K)
 the relaxation times were  $T_1=740~\textrm{ms}$, $T_2=60~\textrm{ms}$, which correspond
 to the unitless values given above for $\omega_{max}/(2 \pi)=32.3$ Hz. For this value of $\omega_{max}$,
 the duration of the intuitive IR sequence is 478 ms, whereas the optimal sequence has a duration of only 202 ms.
 The experimentally measured trajectories of the Bloch vector are also shown in Fig. \ref{fig1} for the optimal
 sequence (filled squares) and the IR sequence (open diamonds). The reasonable match between theory and experiment confirms that the
 complex pulse sequence required for optimization can really be implemented with modern NMR spectrometers.

Figure \ref{fig2} displays the evolution of the optimal solution and of the intuitive one when the maximum amplitude of the control field varies. The ratio between the two control durations $T_{opt}$ and $T_{IR}$ is also plotted as a function of $\omega_{max}/2\pi$. For low values of $\omega_{max}$, the optimal pulse and the IR sequence are very similar and the ratio is close to 1. Note that for $\omega_{max}/2\pi\leq 2.7$, the target state cannot be reached from the initial point so the ratio cannot be defined. We observe a rapid decrease of this ratio when $\omega_{max}$ increases showing the crucial role of the horizontal singular line. The gain tends asymptotically to a constant value when $\omega_{max}\to +\infty$ for fixed values of $T_1$ and $T_2$. In this limit, we can neglect the duration of the different bang controls. Using the relation $\omega_s=\frac{\omega_{max}}{2\pi}u_s=\frac{T_2-2T_1}{2T_1(T_1-T_2)y}$, one obtains by a direct integration of the Bloch equation that
\begin{eqnarray*}
& &T_{opt}\to_{\omega_{max}\to +\infty} \frac{T_2}{2}\log(1-\frac{2}{\alpha T_2})+T_1\log (\frac{2T_1-T_2}{2(T_1-T_2)}),\\
& &T_{IR}\to_{\omega_{max}\to +\infty} T_1\log 2,
\end{eqnarray*}
where $\alpha=\frac{T_2(T_2-2T_1)}{2T_1(T_1-T_2)^2}$, which leads
to a limit ratio of 0.389.\\
\emph{Physical interpretation of the optimal control strategy.} In
the example considered, the role of singular extremals can be
physically interpreted in light of the dissipation effects. We
introduce the polar coordinates $(r,\theta)$ such as
$y=r\cos\theta$ and $z=r\sin\theta$. A straightforward computation
then leads to:
\[\dot{r}=-(\Gamma\cos^2\theta+\gamma\sin^2\theta)r+\gamma\sin\theta,\]  \[d\dot{r}/d\theta=-(\gamma-\Gamma)r\sin(2\theta)+\gamma\cos\theta.\]
One immediately sees that the control field $u$ cannot modify the
radial velocity but only the orthoradial one $\dot{\theta}$. To
reach in minimum time the center of the Bloch ball, the idea is
then at each time to be on the point where $|\dot{r}|$ is maximum
for a fixed value of the radial coordinate $r$. The singular
control $u_s$ defined in Eq. (\ref{sing}) is determined so that
the dynamics stays on the line of maximum variation of the radius
$r$. In other words, this means that the set of solutions of the
equation $d\dot{r}/d\theta=0$ is exactly the set $S$. One deduces
that the strategy of the optimal control can be thought of as
follows. A first bang pulse is applied to the system to reach the
horizontal singular line. The radius $r$ is then optimally reduced
along this curve as long as the control field satisfies the
constraint of the control problem. The local optimality of this
line can be recovered by showing that the points of this set are
associated to maxima of the function $|\dot{r}|(r,\theta)$ for $r$
fixed. When the limit of admissibility is attained, a new bang
pulse is applied to reach the vertical singular line in a region
where this set is optimal. We finally arrive to the target state
along this curve. We recover here a mechanism introduced in
\cite{tannor} for cooling a two-level system.\\
\begin{figure}
\centering
\includegraphics[height=2in]{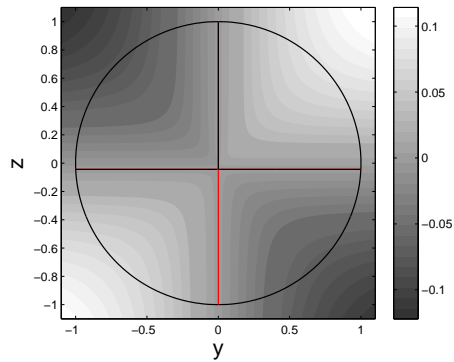}
\caption{\label{fig3} (Color online) Contour plot of the function $d\dot{r}/d\theta$ as a function of $y$ and $z$. The solid lines represent the set of zeros of $d\dot{r}/d\theta$ or the singular set $S$ (see the text). The time-minimum and time-maximum singular lines are plotted respectively in black and red (light gray). The circle is the projection of the Bloch sphere in the plane $(y,z)$.}
\end{figure}
\emph{Conclusion and prospective views.} We hope that this example of singular extremals in the control of a spin 1/2 particle in a dissipative environment will motivate systematic investigations of singular controls in quantum mechanics. The question of the role of singular extremals in more complicated
systems or in quantum computing remains completely open. The next step of this study could be the analysis of the optimal control of coupled spins with bounded fields generalizing thus the different works in this domain with unbounded controls \cite{khaneja}.


\begin{thebibliography}{}

\bibitem{BC} B. Bonnard and M. Chyba, \emph{Singular trajectories and their role in control theory}, Math\'ematiques and Applications, 40, Springer-Verlag, Berlin, 2003.

\bibitem{BS} B. Bonnard and D. Sugny,
SIAM J. on Control and Optimization, \textbf{48}, 1289 (2009);
B. Bonnard, M. Chyba and D. Sugny,
IEEE Transactions on Automatic control, \textbf{54}, 11 (2009); D. Sugny, C. Kontz and H. R. Jauslin,
Phys. Rev. A, \textbf{76}, 023419 (2007); D. Sugny and C. Kontz, Phys. Rev. A, \textbf{77}, 063420 (2008).

\bibitem{boscain2} U. Boscain and G. Charlot, ESAIM: COCV \textbf{10}, 593 (2004).

\bibitem{boscain} U. Boscain and P. Mason,
J. Math. Phys. \textbf{47}, 062101 (2006).

\bibitem{boscainbook} U. Boscain and B. Piccoli, \emph{Optimal syntheses for control systems on 2-D manifolds}, Math\'ematiques and Applications, 43, Springer-Verlag, Berlin, 2004.

\bibitem{bryson} A. E. Bryson and Y.-C. Ho, \emph{Applied Optimal Control}, Taylon and Francis group, New-York-London, 1975.

\bibitem{conolly} S. Conolly, D. Nishimura, A.Macovski,
IEEE Trans. Med. Imag. \textbf{MI-5}, 106
(1986).

\bibitem{alessandro} D. D'Alessandro, IEEE Transactions on Automatic control \textbf{46}, 866 (2001).

\bibitem{khaneja}
N. Khaneja, R. Brockett and S. J. Glaser, Phys. Rev. A, \textbf{63}, 032308 (2001);
%
N. Khaneja, S. J. Glaser and R. Brockett, Phys. Rev. A, \textbf{65}, 032301 (2002);
%
N. Khaneja, T. Reiss, B. Luy, S. J. Glaser,
J. Magn. Reson. \textbf{162}, 311 
(2003).
%
N. Khaneja, B. Luy, S. J. Glaser, 
Proc. Natl. Acad. Sci. USA \textbf{100}, 13162 
 (2003).

\bibitem{spin} M. H. Levitt, \emph{Spin dynamics: Basics of Nuclear Magnetic Resonance}, John Wiley and Sons, New-York-London-Sydney, 2008.

\bibitem{Inv_recov} S. L. Patt, B. D. Sykes,
J. Chem. Phys. \textbf{56}, 3182  
(1972);
 G. M. Bydder, J. V. Hajnal, I. R. Young,
 Clinical Radiology \textbf{53}, 159  
 (1998).

\bibitem{pont} L. S. Pontryagin et al., \emph{The mathematical theory of optimal processes}, John Wiley and Sons, New-York-London, 1962.

\bibitem{rice} S. Rice and M. Zhao, \emph{Optimal control of quantum dynamics}, Wiley, New-York,
2000; M. Shapiro and P. Brumer, \emph{Principles of quantum
control of molecular processes}, Wiley, New-York, 2003.

\bibitem{tannor} S. E. Sklarz, D. J. Tannor and N. Khaneja, Phys. Rev. A, \textbf{69}, 053408 (2004); D. J. Tannor and A. Bartana, J. Phys. Chem. A, \textbf{103}, 10359 (1999).

\bibitem{rabitz} R. Wu, J. Dominy, T.-S. Ho and H. Rabitz, \emph{Singularities of quantum control landscapes}, e-print 0907.2354v1, submitted to J. Math. Phys. (2009).

\end{thebibliography}

\end{document}